\documentclass[11pt]{iopart}

\usepackage{iopams}
\usepackage{graphicx}

\usepackage{dsfont}

\begin{document}

\title{The semiclassical origin of curvature effects in universal spectral statistics}

\author{Daniel Waltner$^1$, Stefan Heusler$^2$, Juan Diego Urbina$^1$, and Klaus Richter$^1$}

\address{$^1$Institut f\"ur Theoretische Physik, Universit\"at Regensburg, 93040 Regensburg, Germany}
\address{$^2$ Institut f\"ur Didaktik der Physik, Universit\"at M\"unster, 48149 M\"unster, Germany}

\begin{abstract}
We consider the energy averaged two-point correlator of spectral determinants and calculate contributions beyond the diagonal approximation using semiclassical methods. Evaluating the contributions originating from pseudo-orbit correlations in the same way as in [S. Heusler {\textit {et al.}}\ 2007 Phys. Rev. Lett. {\textbf{98}}, 044103] we find a discrepancy between the semiclassical and the random matrix theory result. A complementary analysis based on a field-theoretical approach shows that the additional terms occurring in semiclassics are cancelled in field theory by so-called curvature effects. We give the semiclassical interpretation of the curvature effects in terms of contributions from multiple transversals of periodic orbits around shorter periodic orbits and discuss the consistency of our results with previous approaches. 

\end{abstract}


\section{Introduction}
It is currently accepted that progress on the long way to rigorous grounds of the Bohigas-Giannoni-Schmit conjecture \cite{BGS}, stating the emergence of universal fluctuations in the quantum spectra of systems with classical chaotic dynamics, depends on the understanding of correlations among the actions of classical periodic orbits. This idea was pioneered in Berry's seminal paper \cite{berr1}, showing how to incorporate the most basic kind of action correlations into the semiclassical calculation of the spectral form factor. Berry's calculation, referred to as diagonal approximation, considers the contribution from orbit pairs with exactly the same action, \textit{i.e.}\ orbits paired with themselves or their time-reversed counterpart in the case of time reversal symmetry. This calculation reproduces the leading order in a short-time expansion of the form factor obtained from random matrix theory (RMT) (or the leading order in an expansion in the inverse energy difference of the RMT spectral correlator).

\begin{figure*}[htb]
\begin{center}
\includegraphics[height=4cm]{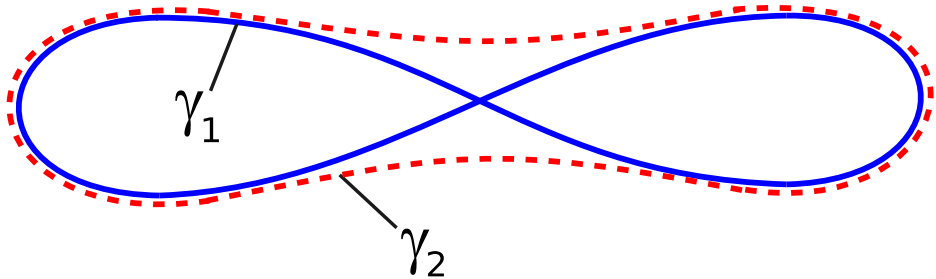}
\caption{Sketch of the orbit correlation mechanism described in \cite{SR}.}
\label{fig1}
\end{center}
\end{figure*}

In seeking classical correlations responsible for the higher order terms Sieber and Richter succeeded \cite{SR}, invoking a subtle shift of the point of view: instead of {\it looking for} pairs of classical orbits with correlated actions, one {\it constructs} them for a given value of the effective Planck's constant. The mechanism underlying the formation of correlated orbit pairs is the exponentially close approach of classical orbits and their time reversed partners everywhere except in the vicinity of a self crossing, as shown in Fig.\ \ref{fig1}. This approach, giving the next to leading order contribution to the RMT form factor in the orthogonal case, was then generalized, leading to the semiclassical calculation of the RMT form factor for times smaller than the Heisenberg time $T_{H}$ \cite{essen1}. Based on the analogy between semiclassics and field theory the form factor for times larger than $T_{H}$ could be obtained semiclassically in \cite{essen2}, although the underlying correlations remained unclear. 

The approaches dealing with times smaller than the Heisenberg time started from the density of states $d(E)$, defined in terms of the spectrum $\{E_n\}$ as
\begin{equation}
\label{den1}
d(E)=\sum_{n=1}^\infty\delta\left(E-E_n \right), 
\end{equation}
which is approximated semiclassically by the Gutzwiller trace formula \cite{gutz},
\begin{equation}
\label{den2}
d^{\rm G}(E)=\bar{d}(E)+\frac{1}{\pi\hbar}\Re\sum_p T_p^{\rm prim}D_p(E) {\rm e}^{iS_p(E)/\hbar}.
\end{equation}
It contains a smooth part, the mean level density $\bar{d}(E)$,  and an oscillatory part expressed as a sum over classical periodic orbits $p$ possessing the classical actions $S_p$, the primitive periods $T_p^{\rm prim}$ and the stability amplitudes $D_p(E)$; for their exact form see \cite{gutz}. The Gutzwiller trace formula incorporates however the drawback that it is divergent, since it contains an infinite sum over an exponentially proliferating number of periodic orbits.  

This problem of the Gutzwiller trace formula renewed the interest in the semiclassical theory of spectral determinants. The quantum spectral determinant is defined as
\begin{equation}
\label{defdel}
\Delta\left(E\right)=\prod_{n=1}^{\infty}\Omega(E,E_{n})(E-E_{n}),
\end{equation}
where $\Omega(E,E_{n})$ is a regularization factor without real zeros. The density of states, considered before, is proportional to the imaginary part of the logarithmic derivative of this determinant. The crudest semiclassical approximation for the spectral determinant is its representation as an infinite product over the classical orbits $p$ occurring in the Gutzwiller trace formula. Equivalently, $\Delta(E)$ can be written as an infinite sum over composite orbits, so called pseudo-orbits $A$ with overall actions and stabilities given, respectively, by the sum and the product of the actions and stabilities of all possible subsets of periodic orbits with $n_{A}$ elements \cite{voros},

\begin{eqnarray}
\label{delG}
\Delta^{{\rm G}}\left(E\right)&=&B(E){\rm e}^{-i\pi\bar{N}(E)}\prod_{p}{\rm exp}\{-D_{p}(E){\rm e}^{iS_{p}(E)/\hbar}\}\nonumber \\
&=&B(E){\rm e}^{-i\pi\bar{N}(E)}\sum_{A}(-1)^{n_{A}}F_{A}(E){\rm e}^{iS_{A}(E)/\hbar}. 
\end{eqnarray}
Here, $\bar{N}(E)$ is the mean number of states up to energy $E$ and $B(E)$ a regularization factor. This form is divergent, as it is only based on the Gutzwiller trace formula. However, by directly imposing that the spectral determinant is real for real energies, a semiclassical theory of the spectral determinant can be constructed. Such an approach was developed by Keating and Berry \cite{berr2} and obtained by Georgeot and Prange using Fredholm theory \cite{fredholm}. Thereby an improved (resurgent) semiclassical approximation to the quantum spectral determinant is obtained. It is given by 
\begin{equation}
\label{eq2}
\Delta^{{\rm R}_{1}}\left(E\right)=2\Re \Delta^{{\rm G}}_{T_{H}/2}\left(E\right),
\end{equation}
where $\Delta^{{\rm G}}_{T_{H}/2}\left(E\right)$ is obtained from $\Delta^{{\rm G}}\left(E\right)$ by including only pseudo-orbits with periods smaller than $T_{H}/2$. Taking the real part assures that $ \Delta^{{\rm R}_{1}}(E)$ is real for real energies. Moreover, truncating the sum over pseudo-orbits makes $\Delta^{{\rm R}_{1}}(E)$ convergent. In this paper we will also consider  
\begin{equation}
\label{eq1}
\Delta^{{\rm R}_{2}}\left(E\right)=2\Re \Delta^{{\rm G}}\left(E\right),
\end{equation}
imposing only the spectral determinant to be semiclassically real.


Starting from the semiclassical approximation $\Delta^{\rm G}(E)$, Ref.\ \cite{essen2} considered a generating function containing in the numerator and in the denominator two spectral determinants at four different energies. By taking derivatives with respect to two energies, identifying afterwards the two energies in the numerator with the two energies in the denominator in two different ways and adding the results of these identifications, they obtain the full RMT form factor. The idea then pursued in Ref.\ \cite{MK} was to use the representation (\ref{eq2}) for the spectral determinant instead of $\Delta^{\rm G}(E)$ and to consider only one identification of the energies. Due to complications from the finite cutoff of the sum defining $\Delta^{{\rm R}_{1}}(E)$, the authors of Ref.\ \cite{MK} used thereby $\Delta^{{\rm R}_{2}}(E)$. 

In this paper we consider the question, how to obtain semiclassically the RMT results for the energy averaged two-point correlator of spectral determinants,

\begin{equation}
\label{cde}
C_{\beta}(e)=\lim_{\bar{d}(E)\to \infty}\frac{\langle \Delta(E+e/2\bar{d}) \Delta(E-e/2\bar{d})\rangle_E}{\langle\Delta(E)^{2}\rangle_E},
\end{equation}
starting from the semiclassical representation $\Delta^{{\rm R}_{2}}(E)$ and using the knowledge about action correlations between orbits and pseudo-orbits. The index $\beta$ in Eq.\ (\ref{cde}) refers to the universality class, \textit{i.e.}\ $\beta=1,2$ for the orthogonal and unitary case, respectively. This problem was first addressed by Kettemann {\it et al.}\ \cite{US} in the context of quantum maps using diagonal approximation. The quantity $C_{\beta=2}(e)$ was then reconsidered in diagonal approximation in Ref.\ \cite{MK} for continuous flows applying the methods of \cite{essen2}. In both cases the calculations show perfect agreement with the universal RMT results for the unitary case

\begin{equation}
\label{RMTU}
C_{\beta=2}^{{\rm RMT}}(e)\propto\frac{\sin \pi e}{\pi e}.
\end{equation}
Here the proportionality sign indicates that we focus on the universal part of $C_{\beta}(e)$ and do not consider the semiclassical calculation of the prefactor. 

In the case of time reversal symmetry, however, Ref.\ \cite{US} could not obtain semiclassically the RMT result
\begin{equation}
\label{RMTO}
C_{\beta=1}^{{\rm RMT}}(e)\propto\left(\frac{\cos \pi e}{(\pi e)^{2}}-\frac{\sin \pi e}{(\pi e)^{3}}\right).
\end{equation}



In this paper we evaluate $C_{\beta}(e)$ beyond the diagonal approximation. Using semiclassical techniques to account for action correlations and its natural extension to pseudo-orbit correlations we first demonstrate in Sec.\ 2 that the pseudo-orbit correlations, shown to cancel in the calculation of the spectral density correlator \cite{essen2}, do {\it not} vanish when one considers the simplest non-rational function of spectral determinants, \textit{i.e.}\ Eq.\ (\ref{cde}). We show furthermore by explicit implementation of $\Delta^{{\rm R}_{1}}(E)$ that the problem, \textit{i.e.}\ the apparent disagreement with RMT, is not solved with the correct truncation of the sum over pseudo-orbits. The third section is devoted to a corresponding field theoretical analysis of $C_\beta(e)$, showing that the additional terms we find in our semiclassical analysis are cancelled in field theory by so-called curvature contributions. In the fourth section we then provide a new type of correlations between pseudo-orbits that allows us to semiclassically recover the RMT result in the unitary and, for the first time, also in the orthogonal case. After a discussion of the consistency of our results with previous results in the fifth section we finally conclude.

\section{Semiclassical calculation}

We semiclassically evaluate the correlator $C_{\beta}(e)$ by means of $\Delta^{{\rm R}_{2}}(E)$. Upon substituting Eqs.\ (\ref{delG}) and (\ref{eq1}) into the numerator of Eq.\ (\ref{cde}), this leads to 

\begin{eqnarray}
\label{eq3}
&&\left\langle \Delta\left(E+\frac{e}{2\bar{d}}\right)\Delta\left( E-\frac{e}{2\bar{d}}\right)\right\rangle_E \nonumber\\ &=&2B^{2}(E)\Re {\rm e}^{-i\pi e}\left\langle\sum_{A,B} F_A F_B^* \left(-1 \right)^{n_A+n_B}  {\rm e}^{\left(i/\hbar \right)\left(S_A-S_B \right)} {\rm e}^{i\pi  e \left( \tau_A+\tau_B\right)  }\right\rangle_E, 
\end{eqnarray}
where $\tau_{A,B}=T_{A,B}/T_{H}$ are the periods of the pseudo-orbits $A, B$ in units of the Heisenberg time $T_{H}=2\pi \hbar \bar{d}$, and where we neglected highly oscillatory terms in $E$. We also expanded the actions of the pseudo-orbits around $E$. In the present context the application of the diagonal approximation is equivalent to taking the actions as uncorrelated random variables, and therefore selecting $A=B$. We then can write the sum over pseudo-orbits as an exponentiated sum over orbits $a$. The Hannay and Ozorio de Almeida sum rule \cite{Han} is afterwards used to replace the sum over orbits, $\sum_a\left|F_a\right|^2(.)$, by $\int_{T_0}^\infty\frac{dT}{T}(.)$, where the minimal period $T_0$ will produce an $e$-independent prefactor that cancels due to the denominator in Eq.\ (\ref{cde}). We finally obtain in diagonal approximation the first term in the squared brackets in

\begin{equation}
\label{ctil}
C_{\beta}(e)\propto-\Re \frac{{\rm e}^{-i\pi e}}{(i\pi e)^{2/\beta}}\left[1+\tilde{C}_{\beta}(e)\right].
\end{equation}
The function $\tilde{C}_{\beta}(e)$ contains all universal effects beyond the diagonal approximation, consisting of both orbit and pseudo-orbit correlations. In the following we compute $\tilde{C}_{\beta}(e)$ to leading order in $1/e$. 

\begin{figure*}[hbt]
\begin{center}
\includegraphics[height=5cm]{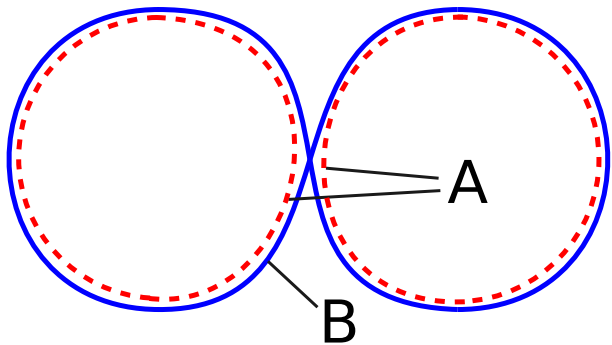}
\caption{Sketch of an ``eight'' diagram, an example of correlated pseudo-orbits \cite{essen2}.}
\label{fig2}
\end{center}
\end{figure*}

In the \textit{unitary case}, the leading order in $1/e$ contribution to $\tilde{C}_{\beta=2}(e)$ comes from the first pseudo-orbit correlations, namely, the ``eight-shaped'' diagrams, see Fig.\ \ref{fig2} \cite{essen2}.
To be more precise, we include in $A$ a pseudo-orbit of order $n$ and in $B$ a pseudo-orbit of order $n+1$ and vice versa. All orbits contained in $A$ and $B$ are assumed to be equal except for the ones forming an ``eight''-orbit. This orbit consists of two components: one long orbit encountering itself at one point and two independent orbits encountering each other near the encounter of the first component. Both components are exponentially close to each other up to deviations in the encounter region. For the evaluation of this contribution we follow Ref.\ \cite{essen2}: $\Delta S=su$ is the action difference of the two components with $s, u$ being the difference of the stable and unstable coordinate in a Poincar\'e surface of section placed in the encounter-region of duration $t_{\mathrm{enc}}=\frac{1}{\lambda}\ln\left( \frac{c^2}{\left|su\right|}\right) $ with a classical constant $c$. Furthermore the weight function counting how often the long orbit of length $T$ has an encounter, is given by $w_T \left(s,u \right)=\frac{T\left( T-2t_\mathrm{{enc}}\right) }{2t_{\mathrm{enc}}\Omega(E)}$, where $\Omega(E)$ is the volume of the energy shell at energy $E$. Employing these expressions we obtain

\begin{eqnarray}
\label{ctilu}
\tilde{C}_{\beta=2}^{{\rm eight}}(e)&=&-2\int_{-c}^c dsdu \int_{2 t_{\mathrm{enc}}}^\infty \frac{dT}{T} w_T\left(s,u \right){\rm e}^{\left(i/\hbar \right)su}{\rm e}^{2\pi ie T/T_{H}}\nonumber \\ &=& -\frac{1}{\pi i e},
\end{eqnarray} 
where the factor 2 in the first line results from the possibility to include the two short orbits in the $A$-\ or $B$-sum in Eq.\ (\ref{eq3}). Using $\tilde{C}_{\beta=2}^{{\rm eight}}(e)$ in Eq.\ (\ref{ctil}) gives a result which is inconsistent with the RMT prediction (\ref{RMTU}). 

In the \textit{orthogonal case} we must consider also the contribution $\tilde{C}_{\beta=1}^{{\infty}}(e)$ from the ``$\infty$-shaped'' orbit pair sketched in Fig.\ \ref{fig1}, apart from the ``eight-shaped'' contribution. Furthermore, in the latter the possible time-reversal operation leaving invariant each of the orbits must be taken into account. The calculations follow the same lines as for the unitary case, and we obtain 

\begin{eqnarray}
\label{ctilo}
\tilde{C}_{\beta=1}^{{\rm eight}}(e)&=&-\frac{4}{i\pi e}, \\
\label{ctilo1}
\tilde{C}_{\beta=1}^{{\infty}}(e)&=&\frac{1}{i\pi e}.
\end{eqnarray}
Again this result contains extra terms resulting from the ``eight-shaped'' diagram when compared with the corresponding RMT result (\ref{RMTO}). We note that the contribution $\tilde{C}_{\beta=1}^{{\infty}}(e)$ alone yields a result consistent with the RMT prediction.

We can conclude that, although the analysis based on the semiclassical representation $\Delta^{{\rm R}_{2}}(E)$ of the spectral determinant produces the same structure as the RMT results (both are naturally ordered in powers in $1/e$), the terms (\ref{ctilu}, \ref{ctilo}) following from semiclassical loop contributions are not consistent with RMT and spoil the agreement found between RMT and the results in \cite{MK,US} based on the diagonal approximation. Moreover, using the same methods we also obtain non vanishing contributions of higher order  in $1/e$ \cite{we1,S2}.

A natural candidate to solve the discrepancy is to use $\Delta^{{\rm R}_{1}}(E)$, Eq.\ (\ref{eq2}), in the semiclassical calculation instead of $\Delta^{{\rm R}_{2}}(E)$. The required truncation of the sum over pseudo-orbits can be exactly incorporated by means of the representation

\begin{equation}
\label{step1}
\Theta\left(T_{H}/2-T_{A}\right)=\frac{1}{2\pi}\int_{0}^{T_{H}/2}dx\int_{-\infty}^{\infty}dk\,{\rm e}^{ik(x-T_{A})}
\end{equation}
for the step function \cite{remark}. Including Eq.\ (\ref{step1}) in the semiclassical calculation presented above, we end up with an integral which must be solved numerically \cite{we1}. The result clearly shows that the extra contributions coming from the ``eight-shaped'' diagram will not be corrected by using the fully resurgent expression (\ref{eq2}) for the semiclassical spectral determinant. Furthermore we rule out that the additional terms result from replacing in $\Delta^{{\rm R}_{1}}(E)$ the smooth step function 
\begin{equation}
\label{step}
\Theta_\sigma\left(T_{H}/2-T_{A}\right)=\frac{1}{\sqrt{\pi}}\int_{\left(T_{A}-T_{H}/2\right)/\sigma}^\infty dx\, {\rm e}^{-x^2},
\end{equation}
suggested in \cite{berr2}, by a sharp cut off. More explicitly, using a Sommerfeld expansion we showed that the smooth cut off only adds to the result with sharp cut off terms proportional to $\sigma^n$ with $n\in\mathds{N}$. These additional terms vanish when $\sigma$ is sent to zero at the end of the calculation, meaning that the two limits $\sigma\rightarrow 0$ and the semiclassical limit $\hbar\rightarrow 0$ commute.

Obviously, the problem requires a better insight into the meaning of the pseudo-orbit correlations. Such an insight is gained through the comparison with a corresponding field-theoretical approach.

\section{Field-Theoretical calculation: the curvature contribution}

In order to better understand the difference between the semiclassical and the RMT results, we calculate $C_{\beta}(e)$ by field-theoretical methods \cite{Ef}. 
Our approach consists of transforming the non-perturbative field-theoretical expression for $C_{\beta}(e)$ into a perturbative  expansion. As we will see, the difference between the semiclassical and the RMT results has a direct correspondence in the field theory. 

We start with the non-perturbative evaluation of the integral following from field-theoretical considerations, which is given by \cite{S2}

\begin{equation}
\label{eq9}
C_{\beta}(e)\propto\frac{1}{4\pi}\int_S d\mu\left(S \right) \, {\rm e}^{-i\frac{\beta \pi e}{4}
{\rm Tr}S}, 
\end{equation}
with $S\in Sp(4)/\left( Sp(2)*Sp(2)\right)$ in the orthogonal and $S\in U(2)/\left(U(1)*U(1)\right) $ in the unitary case. Parameterizing the latter space in terms of angles $\theta$ and $\phi$, we obtain in the unitary case
\begin{eqnarray}
\label{eq9.5}
C_{\beta=2}(e)&\propto&\frac{1}{4\pi}\int_0^{2\pi} d\phi\int_{-1}^1d\cos\theta\, {\rm e}^{-i\pi e\cos\theta} \nonumber \\
&=&\frac{\sin \pi e}{\pi e}
\end{eqnarray}
In a similar way, we obtain in the orthogonal case
\begin{equation}
\label{eq9.6}
C_{\beta=1}(e)\propto\frac{\cos \pi e}{\left( \pi e\right)^2 }-\frac{\sin \pi e}{\left( \pi e\right)^3 },
\end{equation}
in agreement with $C_{\beta}^{\rm RMT}(e)$. A perturbative evaluation of the field-theoretical expressions, in the form of a power-series in $1/e$ that is comparable to semiclassics, is obtained by replacing the integration variables on the sphere by stereographic projection variables in the complex plane. We therefore rewrite $S$ as 
\begin{equation}
\label{eq9.7}
S=\left(\begin{array}{cc}
\mathds{1}_{2/\beta} & 0\\
0 &-\mathds{1}_{2/\beta} \end{array}\right)T\left(\begin{array}{cc}
\mathds{1}_{2/\beta} & 0\\
0 &-\mathds{1}_{2/\beta} \end{array}\right)T^{-1}
\end{equation}
with
\begin{equation}
T=\left(\begin{array}{cc}
\mathds{1}_{2/\beta} & -B^\dagger\\
B &\mathds{1}_{2/\beta} \end{array}\right)
\end{equation}
and $B\in\mathds{C}$ in the unitary and 
\begin{equation}
B=\left(\begin{array}{cc}B_1 & -B_2\\
B_2^* &B_1^* \end{array}\right)
\end{equation}
with $B_1,B_2\in\mathds{C}$ in the orthogonal case. We used here the dagger to indicate the Hermitian conjugate of a matrix and the power $-1$ for its inverse.

The effect of this parameterization is twofold: On the one hand this variable transformation has a non-trivial Jacobian corresponding to a curved measure,
 
\begin{equation}
\label{eq10}
d\mu\left(S \right)=\frac{4\,\left[dB\right] }{\left( 1+\mathcal{B}^2\right)^{4/\beta}}.
\end{equation}
with $\mathcal{B}=\left|B\right|$, $[dB]= dB dB^*$ in the unitary and $\mathcal{B}=\sqrt{\left|B_1\right|^2+\left|B_2\right|^2}$, $[dB]=dB_1 dB_2 dB_1^* dB_2^*$ in the orthogonal case. On the other hand it leads to a particular form of the phase in the exponential

\begin{equation}
\label{eq11}
{\rm Tr} S=\frac{4}{\beta}\left( 1+2\sum_{k=1}^\infty\left(-1\right)^k \mathcal{B}^{2k}\right) .
\end{equation}

In this way and for {\it this particular parameterization} of the supermanifold where the field-theoretical calculation is defined, we can unambiguously identify two kinds of terms in the $1/e$ expansion of the correlator obtained by using in the unitary case $\int_0^\infty dB\left|B\right|^{2k}{\rm e}^{-2\pi i e\left|B\right|^2}\propto\left( 1/e\right)^{k+1}$ and the corresponding relation for $B_1$ and $B_2$ in the orthogonal case. We will call these terms simply phase contributions if the factor in front of the exponential in the last integral results from expanding the right hand side of Eq.\ (\ref{eq11}) and curvature contributions if it results from expanding the right hand side of Eq.\ (\ref{eq10}) in powers of $\mathcal{B}$.

The key point is that we find agreement between the field-theoretical and the semiclassical results (\ref{ctilu}-\ref{ctilo1}), by evaluating Eq.\ (\ref{eq9}) and ignoring the terms originating from the curvature of the measure, \textit{i.e.}\ replacing $(1+\mathcal{B}^{2})^{-4/\beta}$ in Eq.\ (\ref{eq10}) by a constant \cite{we1,S2}. Clearly, what we are missing in the semiclassical calculation are the analogs of the terms that can be derived from the curvature. This means that a new kind of pseudo-orbit correlation is lacking which corresponds to the curvature effects. Moreover, one has to show that this new correlation is consistent with existing semiclassical results showing agreement with RMT. 

It is important to notice that curvature effects did not appear in the case of the spectral generating function in Ref.\ \cite{essen2} because in the corresponding field-theoretical analysis the curvature factor had to be taken there to the $r$-th power with the Replica-index $r$  and was thus vanishing in the Replica-limit $r\rightarrow0$. In practice, this observation can be extended to any measure of the fluctuations which is expressed as an homogeneous rational function of spectral determinants. 

\section{Semiclassical interpretation of curvature effects}

In this section we present the missing pseudo-orbit correlations which provide the curvature effects in the integration over the curved manifold in the field-theoretical approach and thereby consistency with RMT. The essential step is to lift the constraint that one of the short periodic orbits does not surround the other one more than once. This assumption is implicit in the calculation of the ``eight-shaped'' diagrams in Ref.\ \cite{essen2}  when one assumes that both short orbits have periods larger than the encounter time.  

\begin{figure*}[htb]
\begin{center}
\includegraphics[height=5cm]{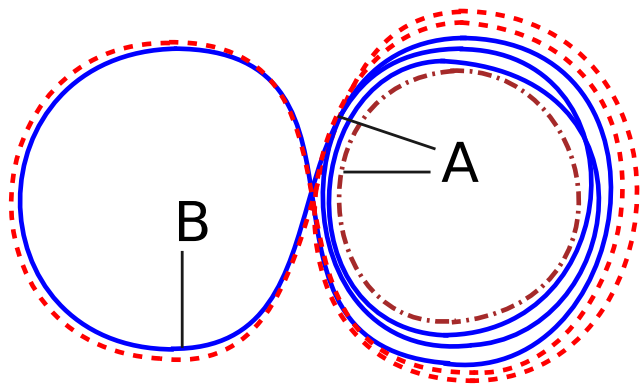}
\caption{Sketch of a pair of pseudo-orbits A, B correlated in a way that one of the orbits (dashed-dotted) has a period smaller than the encounter time. The two long orbits (dashed and full line) differ in the number of transversals around the short periodic orbit by one.}
\label{fig4}
\end{center}
\end{figure*}

To be precise, we consider as components of a pseudo-orbit two periodic orbits (dashed and full line in Fig.\ \ref{fig4}) which include several transversals around a common short periodic orbit and differ in the number of repetitions by one. We then pair the short periodic orbit and the orbit surrounding it $(k-1)$ times with the orbit surrounding it $k$ times. This situation occurs in the case, when the encounter time $t_{\rm{enc}}$ is longer than the shortest of the periodic orbits, \textit{i.e.}\ when the encounter overlaps.

For the evaluation of this contribution, we again need the action difference between the two orbits, obtained as $\Delta S=su$ \cite{Kui} up to a negligible correction and the weight function as a function of $s, u$ that is given by $w_{T,T_p} \left(s,u \right)=\frac{TT_p}{2\Omega(E) t_{\rm{enc}}}$ expressed through the periods of the two short periodic orbits, $T$ and $T_p$. Using these results and applying the sum rule with respect to $T$ and $T_p$, we find for $\tilde{C}_{\beta=2}(e)$ the additional ``curvature'' contribution 

\begin{eqnarray}
\label{eq12}
\tilde{C}_{\beta=2}^{{\rm curv}}(e)&=&-4\int_{-c}^c dsdu \int_{t_{\rm{enc}}}^\infty \frac{dT}{T}\int_0^{t_{\rm{enc}}}\frac{dT_p}{T_p}{\rm e}^{\left(i/\hbar \right)su}{\rm e}^{2\pi ie\left(T+T_p\right)/T_H }w_{T,T_p} \left(s,u \right) \nonumber\\ &=&\frac{1}{i\pi e},
\end{eqnarray} 
thus precisely canceling the undesired term (\ref{ctilu}) coming from the ``eight-shaped'' diagram, and allowing for the full explanation of the RMT result (\ref{RMTU}) in purely semiclassical terms.

In the orthogonal case, one obtains in a similar way

\begin{eqnarray}
\label{eq12.1}
\tilde{C}_{\beta=1}^{{\rm curv}}(e)&=&\frac{4}{i\pi e},
\end{eqnarray} 
canceling again the extra term (\ref{ctilo}) and yielding the universal RMT result (\ref{RMTO}). 

We could thus identify the semiclassical analog of the field theoretical curvature contributions. As in the field-theoretical case the splitting between the two diagrams analyzed in the semiclassical calculation in Sections 2 and 4 is completely arbitrary; the final result, the sum of the two is however always the same. This can be taken into account by considering as reference orbits in the calculation of the weight function always the two short orbits as we did above Eq.\ (\ref{eq12}) and by allowing for an arbitrary length of one of the two periodic orbits as long as the other one is longer than the duration of the encounter. This yields

\begin{eqnarray}
\label{eq12.2}
\hspace*{-2.4cm}\tilde{C}_{\beta=2}^{{\rm eight}}(e)+\tilde{C}_{\beta=2}^{{\rm curv}}(e)&=&-2\int_{-c}^c dsdu \int_{0}^\infty \frac{dT}{T}\int_0^{\infty}\frac{dT_p}{T_p}{\rm e}^{\left(i/\hbar \right)su}{\rm e}^{2\pi ie\left(T+T_p\right)/T_H }w_{T,T_p} \left(s,u \right)\nonumber\\ &&+2\int_{-c}^c dsdu \int_{0}^{t_{\rm{enc}}} \frac{dT}{T}\int_0^{t_{\rm{enc}}} \frac{dT_p}{T_p}{\rm e}^{\left(i/\hbar \right)su}{\rm e}^{2\pi ie\left(T+T_p\right)/T_H }w_{T,T_p} \left(s,u \right)\nonumber \\
&=&0,
\end{eqnarray}
\textit{i.e.}\ the sum of the two contributions calculated in the sections 2 and 4. The splitting of the two contributions in our semiclassical calculation showed however the one-to-one correspondence to the corresponding field theoretical contributions after the stereographic projection.

We want to emphasize here the different behavior of an ``eight'' orbit and the ``$\infty$'' orbit pair in Fig.\ \ref{fig1}: for an ``eight'' orbit with overlapping encounters, one of the short orbits winds around the other and thus yields another nontrivial contribution, that is not yet contained in the diagonal terms. In the case of an ``$\infty$'' orbit pair with overlapping encounters, however, the orbit and its partner transverse the whole orbit in the same direction, which is already contained in the diagonal contribution.

By this calculation we demonstrated that overlapping encounter regions, never considered before in evaluations of energy-averaged quantities neglecting the Ehrenfest-time dependence \cite{Bro}, can lead to non-vanishing contributions even for vanishing Ehrenfest-time. One question, however, has to be addressed at this point: When are effects of encounter overlap important and when can they be neglected? This point will be answered in the following section.

\section{Consistency with former results}

Also in former semiclassical calculations to reproduce RMT-results diagrams involving orbits surrounding a short periodic orbit occurred. Here we discuss the consistency of our results with those approaches.  

\subsection{Approaches involving orbits (but no pseudo-orbits)}

\begin{figure*}
\begin{center}
\includegraphics[height=4cm]
{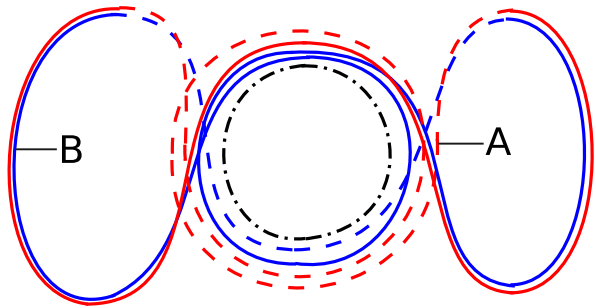}
\caption{Encounter overlap for different encounters}
\label{fig5}
\end{center}
\end{figure*}

In this case an overlap of encounters can only happen for \textit{different} encounters, that means, it is not possible that the ends of \textit{one} encounter come close to each other, because otherwise there exist only disconnected partner orbits.
One simple example for a possible diagram, where a longer orbit surrounds a short periodic orbit is shown in the Fig.\ \ref{fig5}. 

In this case, however, a calculation similar to that for the spectral form factor for finite Ehrenfest-time \cite{Bro} shows that the resulting contribution contains, for an arbitrary time of the surrounded periodic orbit $T_p$, just the configurations of two independent encounters of two orbital parts and one encounter of three parts. Such configurations were already taken into account in \cite{essen1}. In particular, this means that we do not find a different result by also taking into account a configuration where a periodic orbit is surrounded by a longer one many times.

\subsection{Approaches involving pseudo-orbits}

In Ref.\ \cite{essen2}, the contribution from ``eight-shaped'' orbits vanishes also, but for a different reason:  the contributions arising from different possibilities to include the ``eight-shaped'' orbit in the pseudo-orbit sums cancel (giving agreement with RMT). However, this method is not applicable in general, as was shown in this paper. Still this will not pose a problem for measures of the spectral fluctuations based on the density of states, since they involve homogeneous rational functions of spectral determinants. 

\section{Conclusions}
We have shown that the semiclassical approach from Ref.\ \cite{essen2} applied to the two-point correlator of spectral determinants (and we believe this also holds for any measure involving only products of spectral determinants) produces spurious terms coming from pseudo-orbit correlations, spoiling agreement with the universal RMT results. We argued that the appearance of extra terms in the correlator is not solved by the strict use of the resurgent spectral determinant, and that a new additional mechanism is required. This contribution consists of pseudo-orbit correlations represented by encounter diagrams with overlapping encounters. We have shown that these effects are the semiclassical analogue of the curvature contributions in the field theoretical approach, thus providing a semiclassical interpretation of the latter. 

Our results can be extended in various directions: 
First, concerning the study of action correlations between orbits it would be interesting to further analyze the new diagram type with an overlapping encounter identified in this paper: Are there other quantities apart from the product of two spectral determinants, where this correlation is essential to reproduce the RMT-results or do similar correlations lead to further undiscovered contributions? 
Second, as the correlator of two spectral determinants can be used to describe certain features of localization \cite{Ket}, it would be interesting to use the field theoretical calculation presented in Ref.\ \cite{Ket} as a guideline for a semiclassical approach to localization.

\section{Acknowledgements}
We are grateful to M.\ Guti\'errez, S.\ Kettemann, J.\ Kuipers and S.\ M\"uller for helpful discussions and correspondence. DW, JDU and KR acknowledge financial support from DFG (within GRK 638).


\begin{thebibliography}{10}
\bibitem{remark}
A similar  calculation based on a different representation of the step function was attempted in Ref.\ \cite{MK} for the diagonal contribution, but it requires separate analyses of the two limits of large and small energy differences $e$.
\end{thebibliography}
\end{document}